\documentclass[12pt,preprint]{aastex}

\begin{document}
\title{Correlation Properties of Galaxies from the Main Galaxy Sample
of the SDSS Survey}
\author{Anton V.\ Tikhonov}
\affil{ St. Petersburg State University,St.\ Petersburg, Russia,
e-mail: avt@gtn.ru, ti@hotbox.ru}
\email{avt@gtn.ru, ti@hotbox.ru}
\begin{abstract}

The apparatus of correlation gamma function ($\Gamma^*(r)$) is
used to analyze volume-limited samples from the DR4 Main Galaxy
Sample of the SDSS survey with the aim of determining the
characteristic scales of galaxy clustering. Up to $20 h^{-1}$~Mpc
($H_0 = 65$~km~s$^{-1}$Mpc$^{-1}$), the distribution of galaxies
is described by a power-law density-distance dependence,
$\Gamma^*(r) \propto r^{-\gamma}$, with an index $\gamma \approx
1.0$. A change in the state of clustering (a significant deviation
from the power law) was found on a scale of $20-25 h^{-1}$~Mpc.
The distribution of SDSS galaxies becomes homogeneous ($\gamma
\sim 0$) from a scale of $\sim 60 h^{-1}$~Mpc. The dependence of
$\gamma$ on the luminosity of galaxies in volume-limited samples
was obtained. The power-law index $\gamma$ increases with
decreasing absolute magnitude of sample galaxies $M_{abs}$. At
$M_{abs}\sim-21.4$,which corresponds to the characteristic value
$M^*_r$ of the SDSS luminosity function, this dependence exhibits
a break followed by a more rapid increase in $\gamma$.

\end{abstract}


DOI: 10.1134/S1063773706110016

\section{Introduction}
The large body of observational evidence, in particular, the
homogeneity and isotropy of the blackbody cosmic microwave
background radiation, suggests that the distribution of visible
matter is homogeneous on fairly large scales. The homogeneity of
the matter distribution in the Universe is postulated by the
standard model of cosmological evolution (Peebles 1980).  At the
same time, the large-scale structuring of galaxies is traceable on
scales larger than 300 Mpc (the Sloan Great Wall structure of of
the SDSS survey). On small scales, the integrated parameters of
the distribution of galaxies and systems of galaxies exhibit a
power-law decrease in the density of objects as the volume under
consideration increases. This behavior of the density is
occasionally interpreted as fractality, self-similarity of
structures in a certain range of scales. Both the extent of
fractal structures and the possibility of describing the observed
distribution in terms of fractal sets (Davis 1997; Sylos Labini et
al. 1998; McCauley 2002; Barishev and Teerikorpi 2005) are being
debated. If fractality were found on significant scales, then its
formation would have to be explained theoretically and reproduced
in numerical simulations of cosmological evolution. A large extent
of fractal structures (of the order of several hundred Mpc) would
pose serious challenges to the standard theory of galaxy
formation.

To determine the characteristic scales of galaxy clustering, we
use the conditional density function or the correlation gamma
function (Coleman and Pietronero 1992). Based on a sample of rich
Abell clusters, Tikhonov et al. (2000) determined the  scale, $
\sim 100 h^{-1} $Mpc (for the Hubble constant $H_0 =
100$~km~s$^{-1}$Mpc$^{-1}$), from which the distribution clusters
becomes homogeneous using this method. Based on galaxies from the
CfA2 and SSRS2 surveys  and clusters from the APM survey, they
established that the regions in which the density-distance
dependence is well fitted by a single power law are limited by a
scale of $\sim 30 h^{-1} $~Mpc. To test these results, in
particular, to determine the homogeneity scale in the distribution
of galaxies, we use data from the SDSS survey (York et al. 2000),
which is currently the largest survey in sky coverage and depth.
Based on the Luminous Red Galaxy Sample (Eisenstein et al. 2001)
of SDSS galaxies, Hogg et al. (2005) found a homogeneity scale of
$\sim 70 h^{-1} $Mpc (for $H_0 = 100$~km~s$^{-1}$Mpc$^{-1}$). In
this paper, we analyze a sample of galaxies from another part of
the SDSS survey, the Main Galaxy Sample (Strauss et al. 2002).

Another principal objective of the correlation analysis is to
describe the clustering of galaxies on small scales. The galaxy
clustering parameters are known to depend on the luminosity,
morphological type, and color of galaxies (see Zehavi (2005) and
references therein). Bright galaxies exhibit a stronger clustering
than faint galaxies; the differences are more pronounced beginning
from $L^{*}$, the characteristic value of the Schechter luminosity
function for galaxies. A detailed dependence of the clustering on
galaxy properties is required to constrain the parameters of the
theory of galaxy formation, in particular, to ascertain the
initial conditions and to determine the formation conditions of
these properties. This dependence was determined mainly using a
two-point correlation function (see Zehavi (2005) and references
therein). In this paper, we establish the dependence of the degree
of clustering on galaxy luminosity using a correlation gamma
function.

\section{THE DATA}

When fully implemented, the SDSS (Sloan Digital Sky Survey) is
expected to yield spectroscopic redshifts for $\sim 10^6$ galaxies
and $\sim 10^5$ quasars based on photometric data from a sky
region with an area of $10^4$ sq. degrees in the Northern Galactic
Hemisphere in five bands, u, g, r, i,and z, with a limiting
magnitude of $r = 22.5$. The photometric data were used for a
homogeneous selection of objects of various classes (galaxies,
stars, and quasars) to be subsequently observed spectroscopically.
Two types of galaxies were chosen to determine the redshifts from
the list of objects classified as extended ones: galaxies with
Petrosian magnitudes $r < 17.77$ and surface brightnesses higher
than 24m/$\square$" formed the Main Sample (the number of such
objects in the final SDSS version is 900000) and the LRG (Luminous
Red Galaxies) list includes galaxies with very red colors and $r <
19.5$ (the final SDSS version contains 100000 such objects). In
this paper, we analyze data from the fourth SDSS data release DR4
(www.sdss.org) (849920 spectra, including 565715 galaxies and
76484 quasars).

When processing the DR4 data, we first selected a rectangular part
from the region of spectroscopic sky coverage to make the
allowance for the boundary conditions more convenient when
calculating the gamma function and to ensure completeness of the
sample. In the coordinate system of the survey ($\lambda, \eta $)
(with the poles at $\alpha_{2000} = 95^\circ$,$ \delta_{2000} =
0^\circ$ and ами в $\alpha_{2000} = 95^\circ$,$ \delta_{2000} =
0^\circ$ и $\alpha_{2000} = 275^\circ, \delta_{2000} = 0^\circ$;
the point $\lambda=0^\circ, \eta=0^\circ$ corresponds to
$\alpha_{2000} = 185^\circ, \delta_{2000} = 32.5^\circ$, $\eta$
increases northward), the selected region is $-47^\circ < \lambda
< 8^\circ$, $9^\circ < \eta < 36^\circ$. This region contains a
certain number of small gaps in the spectroscopic coverage of the
sky area under consideration. Filling the corresponding regions
with the density equal to the mean density of the galaxy
distribution in the selected region showed these gaps to have
virtually no effect on the result.

The Main Galaxy Sample is an apparent-magnitude-limited survey,
which determined the method for constructing a volume-limited
sample, determining the constraint on the absolute $r$-band
magnitude of sample galaxies $M_{lim} = r_{lim}-25-5lg(R_{max}
(1+z_{max}))-K(z)$), where $r_{lim} = 17.77$ was taken as the
limiting $r$-band magnitude and $K(z)$ is the K-correction. We
chose the sample parameters $z_{max} = 0.16$ and $z_{min} = 0.05$
based on the z-distribution of galaxies (histogram) so as to
select the region of required completeness in redshift and to be
able to fit fairly large spheres into the derived geometrical
boundaries.

We constructed a volume-limited sample containing 18916 galaxies
with $M_{abs} < -21.77$. The mean K correction for SDSS galaxies
in the form $K(z)= 2.3537z^2+0.5735z-0.18437$ (Hikage et al. 2005)
was used to estimate the absolute magnitudes of galaxies. We
recalculated the metric distances from the redshifts using the
Hubble constant $H_0 = 65$~km~s$^{-1}$Mpc$^{-1}$ and the density
parameters .$\Omega_{vacuum} = 0.7$, $\Omega_0 = 0.3$ (see, e.g.,
Hogg 1999).

\section{THE METHOD}
In the classical method of gamma function, the behavior of the
density with distance is calculated in spheres (integral
$\Gamma^*$) and spherical layers (differential $\Gamma$). Any
sample objects the spheres around which are located completely
within the specified geometrical boundaries of the sample can be
taken as the sphere centers.

The differential ($\Gamma$(r))and integral ($\Gamma^*$(r)) gamma
functions are defined by the following formulas:

\begin{equation}
\Gamma(r)=\frac{1}{N}\sum\limits_{i=1}^{N}
  \frac{3}{4\pi r^2 \Delta}\int\limits_{r}^{r+\Delta}n(r_i-r)dr
\end{equation}
\begin{equation}
\Gamma^*(r)=\frac{1}{N}\sum\limits_{i=1}^{N}
  \frac{3}{4\pi r^3}\int\limits_{0}^{r+\Delta}n(r_i-r)dr
\end{equation}
where
$n(\hat{r})=\frac{1}{\tilde{N}}\sum\limits_{i=1}^{\tilde{N}}\delta(r_i-\hat{r})$
is the number density $r$is the working radius, $N$ is the number
of the centers of the spheres involved in the counts, $\tilde{N}$
is the number of objects in the sample, and $\Delta$ is the
thickness of the spherical layer (it is assumed to be small).

The gamma function gives the density of the objects within the
spherical layer $r_0 < r < r_0 + dr$ around each object, i.e., the
density at a given distance from the sample object for the
differential gamma function and the density in spheres of radius
$r_0$ centered on the sample objects for the integral gamma
function (the sphere centers are not included in the counts, i.e.,
the density of the "neighbors" is measured). In the case of
practical implementation, the volume of the spherical layer is
calculated as $\frac{4}{3}\cdot pi \cdot (r^3 - r^3_0)$.

The counts are averaged. If part of a particular sphere goes
beyond the sample boundaries as the working radius $r_0$
increases, then this sphere is excluded from the counts. Thus, as
the working radius increases, only the spheres with the centers
that come increasingly close to the center of the volume covered
by the sample are involved in the counts. The counts are stopped
when the number of remaining spheres $N_{sp} < 10$. The scale on
which the gamma function is used is limited by the radius $R_s$ of
the largest sphere centered on the sample object that fits into
the geometrical boundaries of the sample.

The result is presented on a logarithmic scale as the dependencies
of $log(\Gamma)$ on $log(r_0)$ and of $log(\Gamma^*)$ on
$log(r_0)$. The slope $\gamma$ of the fitting straight line that
is constructed using the selected region of change in $log(r_0)$
determines the correlation dimension of the distribution
(codimension) $D_{corr} = 3 - \gamma$ ($\gamma \geq 0$). If the
distribution is fractal, then $D_{corr} = D$, where $D$ is the
fractal dimension of the distribution.

Generally, a higher slope (a higher $\gamma$ corresponding to a
lower dimension) means a greater mean decrease in density inside
the volume and, consequently, a larger clustering of objects. The
horizontal portions of the plot indicate that the distribution of
objects in the sample is homogeneous on the corresponding scales
($D_{corr} = 3$).

Clearly, the slope in the logarithmic dependence (a good linear
fit) is insufficient to assert that there is fractality on the
corresponding scales (McCauley 1997; Tikhonov 2002). Nevertheless,
the method of gamma function is an appropriate apparatus for
determining the characteristic scales of galaxy clustering
irrespective of whether the fractal interpretation is valid or
not.

\section{CHARACTERISTIC SCALES OF THE GALAXY DISTRIBUTION}
Figure 1 shows the pattern of SDSS galaxy density variations with
distance. We can reliably distinguish the main features of the
galaxy distribution on various scales. Up to $20 h^{-1}$~Mpc, the
gamma function is well fitted by a power law with the index (the
slope in the logarithmic axes) $\gamma = 0.98 \pm 0.02$. On a
scale of ($20-25 h^{-1}$~Mpc, significant deviations from a linear
dependence begin --- the gamma function exhibits a break; the
change in the slope is sharper for $\Gamma$(r), which reflects the
change in the state of clustering more accurately than
$\Gamma^*$(r), since the latter is more inertial. This suggests
that the corresponding changes in the galaxy distribution also
occur on a scale of $20 h^{-1}$~Mpc abruptly. This effect may be
related to the sizes of superclusters. The transition regime of
clustering follows next (the density decreases with distance, but
more slowly than on the power-law portion) and the distribution of
SDSS galaxies becomes homogeneous beginning from a scale of $50-60
h^{-1}$~Mpc. The depth of the sample allows spheres up to $130
h^{-1}$~Mpc (the scale $R_s$) to be fitted in its volume and the
extent of the "homogeneous" portion of the density variation with
distance is considerable.

It follows from our analysis of volume-limited samples with the
angle boundaries $-47^\circ < \lambda < 8^\circ$ and $9^\circ <
\eta < 36^\circ$, different constraints on the radial coordinate
($z_{max}$), and, accordingly, different upper limits on the
absolute magnitudes of the galaxies included in the sample (up to
the sample with the largest sphere that can be fitted in its
boundaries, $\sim 100 h^{-1}$~Mpc) that the form of the gamma
function described above is stable for galaxies with $r$-band
absolute magnitudes $M_{abs}<-21$. Our analysis of samples with
even lower $z_{max}$ (the radius of the largest sphere is smaller
than $100 h^{-1}$~Mpc) showed the stability of the presence and
scale of the break at $20-25 h^{-1}$~Mpc.

To assert that the distribution of galaxies is fractal up to $20
h^{-1}$~Mpc, the power law must be observed in a considerably
larger interval of scales (McCauley 2002), since inhomogeneous
distributions of an arbitrary type can yield a power-law gamma
function in a small interval of scales. At the same time, the
homogeneity of the galaxy distribution on scales larger than $100
h^{-1}$~Iie has been firmly established. The large-scale structure
is characterized by the presence of inhomogeneities with an extent
of more than $100 h^{-1}$~Mpc. For example, the extent of the
Sloan Great Wall structure is believed to be more than $300$~Mpc.
Therefore, the structuring of the Universe is not limited by the
homogeneity scale, but, at the same time, the homogeneity of the
distribution is consistent with the presence of structures on
scales larger than the homogeneity scale (see, e.g., Gaite et al.
1999).

Thus, the conclusions about the pattern of the distribution of
galaxies, clusters, and superclusters reached by Tikhonov et al.
(2000) are confirmed completely.

\section{THE LUMINOSITY DEPENDENCE OF CLUSTERING}

In this paper, we studied the dependence of the slope of the gamma
function before the break on the luminosity range of the galaxies
included in our volume-limited sample. To study the dependence of
the correlation index $\gamma$ over a wide luminosity range, we
chose the redshift boundaries $z_{min} = 0.03$ and $z_{max} =
0.065$. The upper limit on the absolute magnitude of the galaxies
included in our volume-limited sample with these boundaries is
$M_{lim} = -19.58$. We chose the $M_{lim}$ range in which we
analyzed the dependence, $-19.6$ (14703 galaxies) --- $-21.6$
(1381 galaxies), in such a way that the sample was volume-limited
and that a considerable number of galaxies remained in the sample
(the number of objects in the sample decreases with decreasing
$M_{lim}$). In Fig.2, $\gamma$ is plotted against $M_{lim}$. Since
the samples drawn in this way overlap in bright objects, $\gamma$
varies smoothly. The dependence exhibits a break near $M_{lim} =
-20.2$ followed by a more rapid increase in $\gamma$.

To analyze the possible systematic trends related to the change in
the number of objects and in the volume of space covered by the
sample, we analyzed the dependence of $\gamma$ on the chosen far
boundary (the redshift boundary $z_{max}$)at fixed $M_{lim} =
-21.5$ (Fig.3) and the dependence of $\gamma$ on the longitude
sample boundary $\lambda_{max}$ at $\lambda_{min}=-47^\circ$,
$z_{max} = 0.12$, and $M_{lim} = -21.5$ (Fig.4). The index
$\gamma$ increases as $z_{max}$ increases to $0.1$. This may be
explained by the fact that new significant structures fall into
the sample. We can note a slight decrease in $\gamma$ with
increasing $\lambda_{max}$ beginning from $\lambda_{max} =
-10^\circ$ as the volume and the number of galaxies increase.

The dependence of the correlation index $\gamma$ on $M_{abs}$ in
Fig.5 was obtained using a sample with the same geometrical
boundaries as those in Fig.2, but with galaxies taken from
nonoverlapping ranges of absolute magnitudes (the range is 0.2
wide). This increases the statistical significance of the features
in the dependence. The number of objects smoothly decreases from
1861 in the range $-19.6 < M_{abs} < -19.8$ to 555 at $M_{abs} <
-22.0$). The main feature of the dependence in Fig.5 is that
$\gamma$ increases significantly near the characteristic value of
the luminosity function for SDSS galaxies, $M^* = -21.38$ (at $H_0
= 65$~km~s$^{-1}$Mpc$^{-1}$)(Blanton et al. 2003), and above.

\section{CONCLUSIONS}

Thus, as the amount of data obtained as part of the SDSS project
increases, it has become possible to directly test the
cosmological principle. The homogeneity of the galaxy distribution
has passed from a theoretical assumption confirmed by
circumstantial evidence (isotropy of the cosmic microwave
background, homogeneity of the sky distribution of radio and X-ray
sources, etc.) to a strict observational fact, which automatically
closes all the inhomogeneous models of cosmological evolution. The
mean galaxy density is now a directly measurable quantity.

Our results suggest that the slope before the break in the gamma
function on a scale of $\sim 20 h^{-1}$ Mpc (the correlation index
$\gamma$) depends in a complex way on the luminosity and the
region of space in which the measurements are made. This casts
doubt on the possibility of describing the distribution of
galaxies by a single power law (with a single index $\gamma$) on
short scales.

The significant increase in the correlation index $\gamma$ with
decreasing $M_{abs}$ after the break in the luminosity function
for SDSS galaxies is indicative of a close relationship between
the spatial distribution of galaxies and parameters of the
luminosity function for these galaxies. The increase in $\gamma$
with luminosity must be reproduced in model calculations within
the framework of the theory of galaxy formation.

\section{ACKNOWLEDGMENTS}

This work was supported by the Administration
of St. Petersburg (grant no. PD05-1.9-117) and
the "Research and Development in Priority Fields of
Science and Technology" Federal Program (project
no. 02.438.11.7001).

\section{REFERENCES}

1. Y. Barishev and P. Teerikorpi, astro-ph/0505185, (2005).

2. M. R. Blanton, D. W. Hogg, J. Brinkmann, et al., Astrophys. J.
595, 819 (2003); astro-ph/0210215.

3. P. H. Coleman and L. Pietronero, Phys. Rep. 213, 311 (1992).

4. M. Davis, in Proceedings of the Conference: Critical Dialogues
in Cosmology, Princeton, New Jersey, 1996, Ed. N. Turok (World
Sci., Singapore, 1997), p. 13.

5. D. J. Eisenstein, J. Annis, J. E. Gunn, et al., Astron. J. 122,
2267 (2001); astro-ph/0108153.

6. J. Gaite, A. Dominguez, and J. Perez-Mercader, Astrophys. J.
522, L5 (1999).

7. C. Hikage, T.Matsubara,Y.Suto, et al., Publ. Astron. Soc. Jpn.
57, 709 (2005); astro-ph/0506194.

8. D.W.Hogg, D.J.Eisenstein, M. R. Blanton, et al., Astrophys. J.
624, 54 (2005); astro-ph/0411197.

9. D. W. Hogg, astro-ph/9905116 (1999).

10. J. L. McCauley, Physica A 309, 183 (2002); astro-ph/9703046.

11. P. J. E. Peebles, The Large-Scale Structure of the Universe
(Princeton Univ. Press, Princeton, N.J., 1980; Mir, Moscow, 1983).

12. M. A. Strauss, D. H. Weinberg, R. H. Lupton, et al., Astron.
J. 124, 1810 (2002); astro-ph/0206225.

13. F. Sylos Labini, M. Montuori, and L. Pietronero, Phys. Rep.
293, 61 (1998).

14. A. V. Tikhonov, Astrofizika 45, 99 (2002) [Astrophys. 45, 79
(2002)].

15. A. V. Tikhonov, D. I. Makarov, and A. I. Kopylov, Bull. Spwc,
Aastrofiz. Obs. 50, 39 (2000); astro-ph/0106276.

16. D.J.York, J. Adelman, J. E. Anderson, etal.,Astron.
J. 120, 1579 (2000); astro-ph/0006396.

17. I. Zehavi, Z. Zheng, D. Weinberg, et al., Astrophys. J. 630, 1
(2005); astro-ph/0408569.

\newpage
\begin{figure}
\centerline{
\includegraphics[]{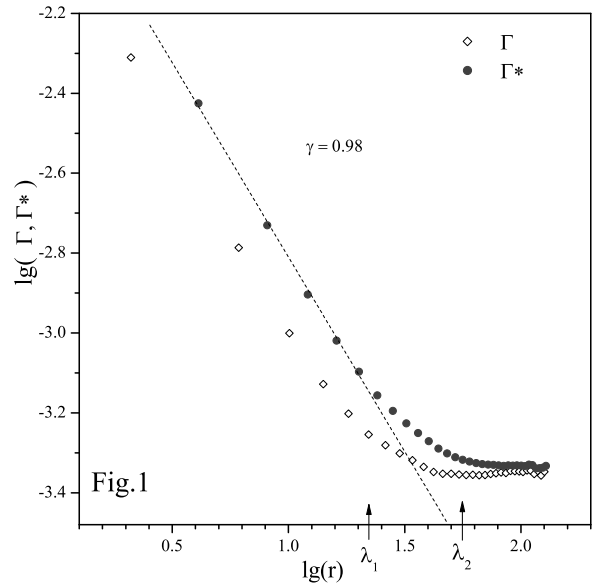}
}
\figcaption{Gamma function of a volume-limited sample of
galaxies from the SDSS Main Galaxy Sample with $z_{min} = 0.05$,
$z_{max} = 0.16$, $M_{abs} < -21.77$, $N = 18916$ is the number of
galaxies in the sample, $R_s = 132$~Mpc, $\lambda_1$ is the break
scale, and $\lambda_2$ is the homogeneity scale. }
\end{figure}

\begin{figure}
\centerline{
\includegraphics[]{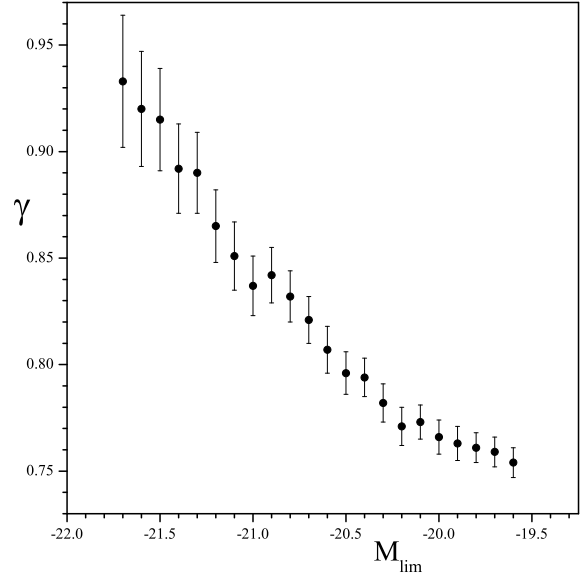}
}
\figcaption{Correlation index $\gamma$ vs. limit on the absolute
magnitude of the galaxies, $M_{lim}$, included in the sample with
$0.03<z<0.065$, $-47^\circ < \lambda < 8^\circ$, $9^\circ < \eta <
36^\circ$.}
\end{figure}

\begin{figure}
\centerline{
\includegraphics[]{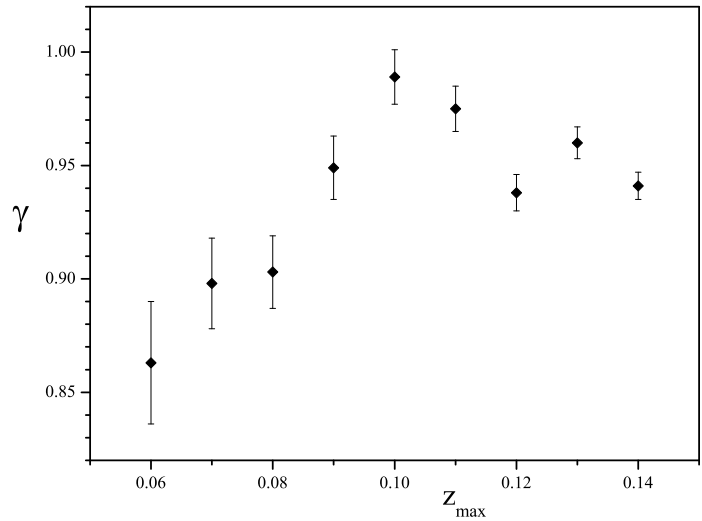}
}
\figcaption{Correlation index $\gamma$ as a function of the far
sample boundary $z_{max}$ at $z_{min} = 0.03$ and $M_{lim} =
-21.5$.}
\end{figure}

\begin{figure}
\centerline{
\includegraphics[]{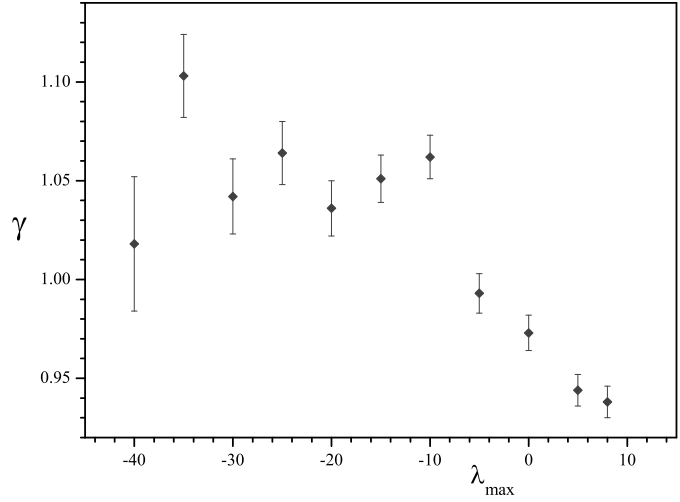}
}
\figcaption{Correlation index $\gamma$ as a function of the
longitude sample boundary $\lambda_{max}$ at
$\lambda_{min}=-47^\circ$, $z_{max} = 0.12$ and $M_{lim} = -21.5$.
}
\end{figure}

\begin{figure}
\centerline{
\includegraphics[]{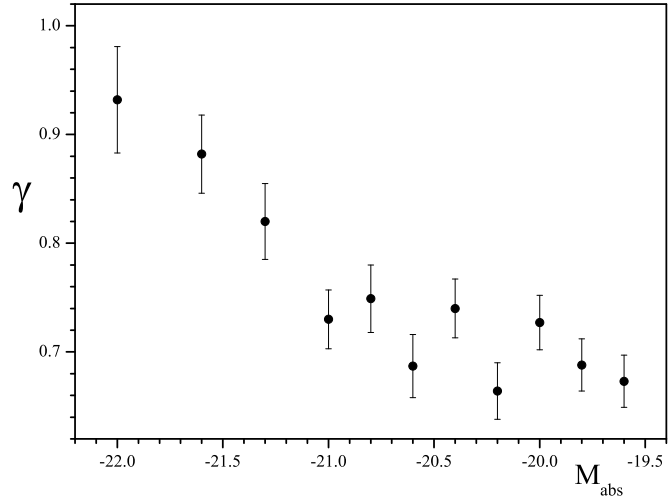}
} \figcaption{Correlation index $\gamma$ vs. $M_{abs}$ range for
nonoverlapping ranges of absolute magnitudes at $0.03<z<0.065$,
$-47^\circ < \lambda < 8^\circ$, $9^\circ < \eta < 36^\circ$. The
value of $\gamma$ corresponds to the upper boundary of the
absolute magnitude interval and the next value corresponds to the
lower boundary in the direction of decreasing $M_{abs}$. }
\end{figure}

\end{document}